\documentclass[aps,preprint]{revtex4}
\usepackage{epsfig}

\begin{document}
\author{Yu Jiang$^{1}$, Yuan-mei Shi$^{1}$, Hua Li$^{1}$, Wei-min Sun$^{1,2}$ and Hong-shi Zong$^{1,2}$}
\address{$^{1}$ Department of Physics, Nanjing University, Nanjing 210093, China}
\address{$^{2}$ Joint Center for Particle, Nuclear Physics and Cosmology, Nanjing 210093, China}
\title{The Calculation of $f_\pi$ and $m_\pi$  at Finite Chemical Potential}
\begin{abstract}
Based on the previous work in [Y. Jiang, Y.M. Shi, H.T. Feng, W.M.
Sun and H.S. Zong, Phys. Rev. C {\bf 78}, 025214 (2008)] on the quark-meson vertex and pion properties at finite quark chemical potential, we provide an analytical analysis of the weak decay constant of the pion ($f_\pi[\mu]$) and the pion mass ($m_\pi[\mu]$) at finite quark chemical potential using the model quark propagator proposed in [R. Alkofer, W. Detmold, C.S. Fischer and P. Maris, Phys. Rev. D {\bf 70}, 014014 (2004)]. It is found that when $\mu$ is below a threshold value $\mu_0$ 
(which equals $0.350~\mathrm{GeV}$, $0.377~\mathrm{GeV}$ and $0.341~\mathrm{GeV}$, for the $\mathrm{2CC}$, $\mathrm{1R1CC}$ and $\mathrm{3R}$ parametrizations of the model quark propagator, respectively.), $f_\pi[\mu]$ and $m_\pi[\mu]$ are kept unchanged from their vacuum values. The value of $\mu_0$ is intimately connected with the pole distribution of the model quark propagator and is found to coincide with the threshold value below which the quark-number density vanishes identically. Numerical calculations show that when $\mu$ becomes larger than $\mu_0$, $f_\pi[\mu]$ exhibits a sharp decrease whereas $m_\pi[\mu]$ exhibits a sharp increase. A comparison is given between the results obtained in this paper and those obtained in previous literatures.

\bigskip

Key-words: weak decay constant of pion, pion mass, finite quark chemical
potential

\bigskip

E-mail: zonghs@chenwang.nju.edu.cn.

\bigskip

PACS Numbers: 11.10.Wx, 11.10.St, 11.15.Tk, 14.40.Aq

\end{abstract}
\maketitle

The in-medium modification of the properties of the pion is of fundamental interest in hadron physics. The pion is identified as a Goldstone boson arising from the spontaneous breakdown of chiral symmetry which is essential for describing low-energy hadronic phenomena. Since chiral symmetry is expected to be restored at high enough density, the change of pion properties in medium will provide crucial information on the restoration of chiral symmetry. Among these, the weak decay constant of the pion $f_{\pi}$ and the pion mass $m_{\pi}$ are the two most important quantities, since they are closely related to the spontaneous breakdown of chiral symmetry of Quantum Chromodynamics (QCD). Unfortunately, so far it has not been possibile to obtain detailed information about modification of pion properties in medium directly from QCD. In this situation, different models have been used to study this sort of problems \cite{Delorme,Kirchbach,Kaiser,Meissner,Kim,Mallik,Nam,Maris, Bender,Bender1}. Just as was pointed out in Ref. \cite{Maris1}, the pion has a dual role: it can be identified as a quark-antiquark bound state as well as a Goldstone boson arising from the spontaneous breakdown of chiral symmetry. From the point of view that the pion can be regarded as a quark-antiquark bound state, the full dynamical information of the pion is contained in
the corresponding Bethe-Salpeter Amplitude (BSA): $\Gamma_\pi(k,p)$
($k$ is the relative and $p$ the total momentum of the
quark-antiquark pair), which is the one-particle-irreducible,
fully-amputated quark-meson vertex. The Dyson-Schwinger equations (DSEs) of QCD provide a nonperturbative, continuum framework for
analyzing such quark-meson vertices directly \cite{Maris1,DSE1,DSE2,DSE3,DSE4}. The aim of this paper is to study the change of $f_\pi$ and $m_\pi$ with quark chemical potential $\mu$ in the framework of this nonperturbative QCD model. 

The DSEs of QCD have been used
extensively at zero temperature and zero quark chemical potential
to extract hadronic observables \cite{DSE1,DSE2,DSE3,DSE4}.
However, this is very difficult at finite quark
chemical potential due to the fact that the number of independent Lorentz structures of the quark-meson vertex at finite $\mu$ is much larger than that of the corresponding one at $\mu=0$.  In Ref. \cite{fpi1}, using the method of studying the dressed quark propagator at finite $\mu$ given in Ref. \cite{Zong05}, the authors have given a new approach for tackling this problem. Based on the rainbow-ladder
approximation of the DSEs and the assumption of analyticity of the
quark-meson vertex in the neighborhood of $\mu=0$ and neglecting the
$\mu$-dependence of the dressed gluon propagator, the authors show
that the general quark-meson vertex at finite $\mu$ can be obtained
from the corresponding one at $\mu=0$ by a shift of variable:
$\Gamma[\mu](k,p)=\Gamma(\tilde{k},p)$, where
$\tilde{k}=(\vec{k},k_4+i\mu)$. From this result the authors of
Ref. \cite{fpi1} numerically calculated $f_\pi[\mu]$ and $m_\pi[\mu]$ for $\mu<300~\mathrm{MeV}$. It is found that $f_\pi[\mu]$ increases slowly (with an increase of less than about $0.01\%$) and $m_\pi[\mu]$ falls slowly (with a decrease of less than about $0.06\%$) with increasing $\mu$. Numerically the change of $f_\pi[\mu]$ and $m_\pi[\mu]$ is so small that one can think $f_\pi[\mu]$ and $m_\pi[\mu]$ does not change with $\mu$ for $\mu<300~\mathrm{MeV}$ within numerical errors. One of our motivations for this work is to explore the mathematical reason behind this. 
Based on the work in \cite{fpi1}, in this paper we provide an analytic analysis of $f_\pi[\mu]$ and $m_\pi[\mu]$.
It is found that when $\mu$ is below a critical value $\mu_0$, $f_\pi[\mu]$ and $m_\pi[\mu]$ are kept unchanged from their vacuum values. Moreover, numerical calculations show that when $\mu$ becomes larger than $\mu_0$, $f_\pi[\mu]$ exhibits a sharp decrease whereas $m_\pi[\mu]$ exhibits a sharp increase. 

According to Ref. \cite{fpi1}, the pion decay constant at finite $\mu$ can be
expressed as the following
\begin{equation}
\label{fpi}
\delta^{ij}f_{\pi}[\mu]p_{\nu}=\int_q\,\mbox{tr}\left[\frac{\tau^i}{2}
\gamma_5\gamma_\nu
S(\tilde{q}_+)\Gamma_\pi^j(\tilde{q};p)S(\tilde{q}_-)\right],
\end{equation}
where $S(q)$ is the full dressed quark propagator, $\tilde{q}_\pm=\tilde{q}\pm p/2$,
$\tilde{q}=(\vec{q},q_4+i\mu)$, $\frac{\tau^i}{2}$ are the flavor $SU(2)$ generators and
$\int_q\equiv\int\,d^4q/(2\pi)^4$. In the present paper we will not
write the renormalisation constants explicitly because one would find
that in the final result the renormalisation constants cancel each
other. In fact, Eq. (\ref{fpi}) is the expression of $f_\pi$ which is
independent of the renormalisation point and the regularisation
mass-scale \cite{Maris1}.

The integral of the right-hand-side of Eq. (\ref{fpi}) can be
rewritten as:
\begin{equation}
\label{integral}
\int_q\equiv\int\,\frac{d^4q}{(2\pi)^4}\equiv\int\,\frac{d^4\tilde{q}}{(2\pi)^4}.
\end{equation}
Contracting both sides of Eq. (\ref{fpi}) with $p_\nu$ and using Eq.
(\ref{integral}), we obtain the following:
\begin{eqnarray}
\delta^{ij}f_{\pi}[\mu]&=&\frac{1}{p^2}\int\limits_{-\infty}^{+\infty}\,
\frac{d^3\vec{q}}{(2\pi)^3}\int\limits _{-\infty+i\mu}^
{+\infty+i\mu}\,\frac{dq_4}{(2\pi)}
\mbox{tr}\left[\frac{\tau^i}{2}\gamma_5{\not\!p} S(q_+)\Gamma_\pi^j(q;p)S(q_-)\right]\nonumber\\
\label{fpi1}&=&\frac{1}{p^2}\int\limits_{-\infty}^{+\infty}\,\frac{d^3\vec{q}}
{(2\pi)^3}\int_{C_1}\frac{dq_4}{(2\pi)}
\mbox{tr}\left[\frac{\tau^i}{2}\gamma_5{\not\!p}
S(q_+)\Gamma_\pi^j(q;p)S(q_-)\right]
\end{eqnarray}
where the integration path $C_1$ is depicted in Fig. 1.
\begin{center}
\epsfig{file=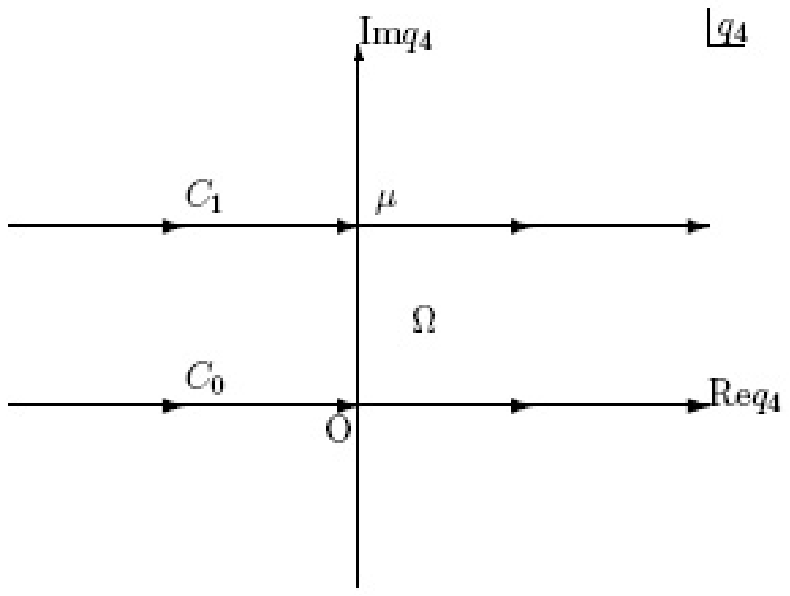, width=9cm}

\vspace{-0.4cm}

{ FIG.1. The integration path in the complex $q_4$ plane.}
\end{center}

Let us use $z_n=\chi_n+i\omega_n ~(\omega_n>0), n=1,2 \cdots$ to denote
the poles of the function
\begin{equation}
\label{function}F^{ij}(q_4)\equiv
\frac{1}{p^2}\mbox{tr}\left[\frac{\tau^i}{2}\gamma_5{\not\!p}
S(q_+)\Gamma_\pi^j(q;p)S(q_-)\right]
\end{equation}
located in the upper half complex $q_4$ plane.
According to Cauchy's theorem we obtain the following from Eq.
(\ref{fpi1}):
\begin{eqnarray}
\delta^{ij}f_{\pi}[\mu]&=&\frac{1}{p^2}\int\limits_{-\infty}^{
+\infty}\,\frac{d^3\vec{q}}{(2\pi)^3} \int_{C_1}\frac{dq_4}{(2\pi)}
\mbox{tr}\left[\frac{\tau^i}{2}\gamma_5{\not\!p}
S(q_+)\Gamma_\pi^j(q;p)S(q_-)\right]\nonumber\\
&=&\frac{1}{p^2}\int\limits_{-\infty}^{+\infty}\,\frac{d^3\vec{q}}
{(2\pi)^3}\int_{C_0}\frac{dq_4}{(2\pi)}
\mbox{tr}\left[\frac{\tau^i}{2}\gamma_5{\not\!p}
S(q_+)\Gamma_\pi^j(q;p)S(q_-)\right]\nonumber\\
&&-i\int\limits_{-\infty}^{+\infty}\,\frac{d^3\vec{q}}{(2\pi)^3}
\sum_n\theta(\mu-\omega_n)\mbox{Res}\{F^{ij}(z);z_n\}\nonumber\\
\label{fpi2}&=&\delta^{ij}f_{\pi}
-i\int\limits_{-\infty}^{+\infty}\,\frac{d^3\vec{q}}{(2\pi)^3}
\sum_n\theta(\mu-\omega_n)\mbox{Res}\{F^{ij}(z);z_n\}.
\end{eqnarray}
From Eq. (\ref{fpi2}) it is easily seen that when $\mu<min\{\omega_n\}$, the function
$F^{ij}(q_4)$ has no pole in the region $\Omega$ (the region
enclosed by $C_1$  and $C_0$, see Fig. 1) and therefore
$f_\pi[\mu]=f_\pi$, which means that for small enough $\mu$ the pion
decay constant should be independent of $\mu$. Of course, when
$\mu>min\{\omega_n\}$ the pion decay constant can have an explicit
$\mu$-dependence.

In the chiral limit, expanding the trace term of the right-hand-side
of Eq. (\ref{function}) to $\mathcal{O}(p^2)$ near $p=0$ \cite{DSE1},  we have the
following:
\begin{eqnarray}
F^{ij}(q_4)&=&\frac{1}{p^2}\mbox{tr}\left\{\frac{\tau^i}{2}\gamma_5{\not\!p}\bigg[
S+\frac{1}{2}p\cdot\partial S\bigg]\bigg[\Gamma_\pi^j(q,0)+\mathcal
{O}(p)\gamma_5\bigg]\bigg[S-\frac{1}{2}p\cdot\partial
S\bigg]\right\},
\end{eqnarray}
where we have adopted the approximation \cite{DSE1}
\begin{equation}
\Gamma_\pi^j(q,p)=\Gamma_\pi^j(q,0)+\mathcal
{O}(p)\gamma_5.\label{app1}
\end{equation}
With this approximation $\Gamma_\pi^j(q,0)$ can
be expressed as \cite{DSE1,Frank96}
\begin{equation}
\label{GammaPi}
\Gamma_\pi^j(q,0)=\tau^j\gamma_5\cdot\frac{iB(q^2)}{f_\pi},
\end{equation}
where $B(q^2)$ is the scalar part of $S^{-1}(q)$.
Noticing that $\mbox{tr}\big[\gamma_5{\not\!p}S\gamma_5 S\big]=0$, we
obtain the following:
\begin{eqnarray}
\label{function2}
F^{ij}(q_4)&=&\frac{1}{p^2}\mbox{tr}\left\{\frac{\tau^i}{4}\gamma_5{\not\!p}\bigg
[p\cdot\partial S\Gamma_\pi^j(q,0)S-S\Gamma_\pi^j(q,0)p\cdot\partial
S\bigg]\right\}+\mathcal {O}(p).
\end{eqnarray}
 Substituting Eq.
(\ref{GammaPi}) into Eq. (\ref{function2}) and using
$\mbox{tr}(\tau^i\tau^j)=2\delta^{ij}$, we obtain
\begin{eqnarray}
\label{function3}
F^{ij}(q_4)&\simeq&\frac{1}{2p^2}\delta^{ij}\frac{iB(q^2)}{f_\pi}\mbox{tr}
\left\{\gamma_5{\not\!p}\bigg [p\cdot\partial
S\gamma_5S-S\gamma_5p\cdot\partial S\bigg]\right\}.
\end{eqnarray}
Adopting the following expression of $S(q)$
\begin{equation}
\label{quarkP}S(q)=\frac{1}{i{\not\!q}A(q^2)+B(q^2)}
=-i{\not\!q}\sigma_v(q^2)+\sigma_s(q^2),
\end{equation}
we obtain
\begin{eqnarray}
\label{function5}F^{ij}(q_4)
&\simeq&\frac{1}{2}\delta^{ij}\frac{1}{f_\pi}\frac{8\sigma_s}
{\sigma_v^2q^2+\sigma_s^2}\left[ \sigma_s\sigma_v+\frac{2(p\cdot
q)^2}{p^2}(\sigma_s\sigma_v^\prime-\sigma_s^\prime\sigma_v)\right]\\
\label{function6}&=&\delta^{ij}\frac{1}{f_\pi}F(q_4),
\end{eqnarray}
where $^\prime$ means $d/dq^2$ and
\begin{eqnarray}
\label{F1}F(q_4)&\equiv& \frac{4\sigma_s}
{\sigma_v^2q^2+\sigma_s^2}\left[ \sigma_s\sigma_v+2\frac{(p\cdot
q)^2}{p^2}(\sigma_s\sigma_v^\prime-\sigma_s^\prime\sigma_v)\right].
\end{eqnarray}
Then Eq. (\ref{fpi2}) can be written as
\begin{equation}
\label{fpi3}f_\pi[\mu]\simeq f_\pi
-\frac{i}{f_\pi}\int\limits_{-\infty}^{+\infty}\,\frac{d^3\vec{q}}{(2\pi)^3}
\sum_n\theta(\mu-\omega_n)\mbox{Res}\{F(z);z_n\}.
\end{equation}

To determine the pole distribution of function $F(q_4)$, we should
first specify the form of the dressed quark propagator. Here, as
in Refs. \cite{fpi1,con1} we adopt the following propagator proposed
in Ref. \cite{Alkofer04}:
\begin{equation}
\label{QuarkP1}
S(q)=\sum_{j=1}^{n_P}\left(\frac{r_j}{i{\not\!q}+a_j+ib_j}+\frac{r_j}{i{\not
\!q}+a_j-ib_j}\right).
\end{equation}
The propagator of this form has $n_P$ pairs of complex conjugate
poles located at $a_j\pm ib_j$. When some $b_j$ is set to zero, the
pair of complex conjugate poles degenerates to a real pole. The
restrictions of the parameters $r_j$, $a_j$ and $b_j$ in the chiral
limit are \cite{Alkofer04}
\begin{eqnarray}
\label{restrictionofr} &&\sum_{j=1}^{n_P}\,r_j=\frac{1}{2},\\
\label{restrictionofra}&&\sum_{j=1}^{n_P}\,r_ja_j=0.
\end{eqnarray}
If we are not in the chiral limit, the right hand side of Eq.
(\ref{restrictionofra}) should be replaced by the current quark
mass. The value of these parameters are shown in Table I, where 2CC,
1R1CC and 3R stand for three meromorphic forms of the quark
propagator, respectively: two pairs of complex conjugate poles, one real pole and
one pair of complex conjugate poles, three real poles.

\begin{center}
\begin{minipage}{16cm}
\begin{center}
{\scriptsize Table I. The parameters used in the calculation of $F(q_4)$ and
$f_\pi$. These parameters are taken directly from Ref.
\cite{Alkofer04}.

\vspace{0.1cm}

\begin{tabular*}{16cm}{l@{\extracolsep{\fill}}*{8}{c}} \hline\hline
Parameterization&$r_1$&$a_1$ (GeV)&$b_1$ (GeV)&$r_2$&$a_2$
(GeV)&$b_2$ (GeV)&$r_3$&$a_3$
(GeV)\\
\hline 2CC&0.360&0.351&0.08&0.140&-0.899&0.463&-&-\\
\hline 1R1CC&0.354&0.377&-&0.146&-0.91&0.45&-&-\\
\hline 3R&0.365&0.341&-&1.2&-1.31&-&-1.06&-1.40\\
\hline\hline
\end{tabular*}}
\end{center}
\end{minipage}
\end{center}

Without losing generality we assume $p_\nu=(\vec{0},p)$ (i.e.
the pion is at rest) and write
\begin{equation}
\frac{(p\cdot q)^2}{p^2}=\frac{q_4^2p^2}{p^2}=q_4^2.
\end{equation}

Now let us calculate $F(q_4)$. With the quark propagator given in
Eq. (\ref{QuarkP1}) we can obtain
\begin{eqnarray}
F(q_4)&=&\frac{\Xi(q_4^2)}
{\prod\limits_j[q^2+(a_j+ib_j)^2]^2[q^2+(a_j-ib_j)^2]^2
\prod\limits_k(q^2+\eta_k^2)},\label{F4}
\end{eqnarray}
where $\Xi$ is a polynomial of $q_4^2$ (for the detailed calculation
of $F(q_4)$, $\Xi$ and $\eta_k$, see the Appendix). The values of
$\eta_k$ are shown in Table II ($\eta_k$ are ordered from small to
large according to their real part).

\begin{center}
\begin{minipage}{16cm}
\begin{center}
{\scriptsize Table II. The calculated values of $\eta_k$.

\vspace{0.1cm}

\begin{tabular*}{16cm}{l@{\extracolsep{\fill}}*{5}{c}} \hline\hline
Parameterization&$\eta_1$ (GeV)&$\eta_2$ (GeV)&$\eta_3$
(GeV)&$\eta_4$ (GeV)&$\eta_5$ (GeV)\\
\hline 2CC&0.350&0.723-0.351i&0.723+0.351i&-&-\\
\hline 1R1CC&0.377&0.723-0.328i&0.723+0.328i&-&-\\
\hline 3R&0.341&0.617&1.31&1.40&1.849\\
\hline\hline
\end{tabular*}}
\end{center}
\end{minipage}
\end{center}
Here it should be noticed that when some $b_j=0$ (the quark propagator has
a real pole), some $\eta_k$ must exactly equal the corresponding
$|a_j|$ (see the Appendix). For 1R1CC case, $b_1=0$ and $\eta_1=|a_1|$.
For 3R case, all $b_j=0$ and $\eta_1=|a_1|$, $\eta_3=|a_2|$,
$\eta_4=|a_3|$. For 2CC case, because $b_1=0.08~\mbox{GeV}$ is very
close to zero, the value of $\eta_1$ is very close to $a_1$.

Because $q^2=q_4^2+\vec{q}^2$, according to Eq. (\ref{F4}) the
poles of $F(q_4)$: $z_n=\chi_n+i\omega_n$ are decided by the
following equation
\begin{eqnarray}
(\chi_n+i\omega_n)^2+\vec{q}^2+(\xi_{nR}+i\xi_{nI})^2=0\label{poles5},
\end{eqnarray}
where $\xi_{nR}$ and $\xi_{nI}$ are the real and imaginary part of
$\eta_k$ or $a_j\pm ib_j$. One can easily find
\begin{eqnarray}
\label{poles4}\omega_n&=&\sqrt{\frac{(\vec{q}^2+\xi_{nR}^2-\xi_{nI}^2)+
\sqrt{(\vec{q}^2+\xi_{nR}^2-\xi_{nI}^2)^2+4\xi_{nR}^2\xi_{nI}^2}}{2}}\\
\chi_n&=&-\frac{\xi_{nR}\xi_{nI}}{\omega_n}.
\end{eqnarray}
From Eq. (\ref{poles4}) we find that for $\mu<|\xi_{nR}|$ the
corresponding $\omega_n$ is always larger than $\mu$, irrespective
of $\vec{q}$. For $\mu>|\xi_{nR}|$, $\omega_n<\mu$ when
$\vec{q}^2<\mu^2-(\xi_{nR}^2\xi_{nI}^2/\mu^2)-\xi_{nR}^2+\xi_{nI}^2$,
and $\omega_n>\mu$ when
$\vec{q}^2>\mu^2-(\xi_{nR}^2\xi_{nI}^2/\mu^2)-\xi_{nR}^2+\xi_{nI}^2$.
Therefore Eq. (\ref{fpi3}) can be written as
\begin{equation}
\label{fpi4}f_\pi[\mu]=f_\pi
-\frac{i}{2\pi^2f_\pi}\sum_n\theta(\mu-|\xi_{nR}|)
\int\limits_{0}^{\Lambda_n(\mu)}\,d|\vec{q}|
\,\vec{q}^2\mbox{Res}\{F(z);z_n\},
\end{equation}
where
\begin{eqnarray}
\Lambda_n(\mu)&=&\sqrt{\mu^2-(\xi_{nR}^2\xi_{nI}^2/\mu^2)
-\xi_{nR}^2+\xi_{nI}^2}\,.
\end{eqnarray}
From Eq. (\ref{fpi4}) and the values of $a_j$, $b_j$ and $\eta_k$ in
Table I and II we find that when $\mu$ is below some threshold value
$\mu_0$, the pion decay constant at finite chemical potential
$f_\pi[\mu]$ is kept unchanged from its vacuum value. The threshold
value $\mu_0$, which equals the minimum of the real part of $a_j\pm
ib_j$ and $\eta_k$, is shown in Table III.
\begin{center}
\begin{minipage}{16cm}
\begin{center}
{\scriptsize Table III. The calculated values of $\mu_0$.

\vspace{0.1cm}

\begin{tabular*}{8cm}{l@{\extracolsep{\fill}}*{2}{c}} \hline\hline
Parameterization&$\mu_0$ (GeV)\\
\hline 2CC&0.350\\
\hline 1R1CC&0.377\\
\hline 3R&0.341\\
\hline\hline
\end{tabular*}}
\end{center}
\end{minipage}
\end{center}
Here we note that in Ref. \cite{Zong08} it is found that when $\mu$ is below the same threshold value $\mu_0$, the quark-number density vanishes identically. Namely, $\mu=\mu_0$ is a singularity which separates two regions with different quark-number densities. In fact, in Ref. \cite{Halasz98}, based on a universal argument, it is pointed out that the existence of some singularity at the point $\mu=\mu_0$ and $T=0$ is a robust and model-independent prediction. Below $\mu=\mu_0$, the QCD system at finite $\mu$ remains in the vacuum (ground state) of QCD at $\mu=0$, so the properties of the Goldstone boson excited from this vacuum does not change with $\mu$. Thus the result that $f_\pi[\mu]$ is kept unchanged from its vacuum value is just to be expected. Here it
should also be noticed that in our method the value of $\mu_0$ is
intimately connected with the pole distribution of the quark
propagator.

In Ref. \cite{con1}, with the same quark propagator the authors find
that the quark condensate at finite chemical potential is kept
unchanged from its vacuum value when $\mu<\mu_0$. From the
Gell-Mann-Oakes-Renner relation $f_\pi^2[\mu]
m_\pi^2[\mu]=2m\langle\bar{q}q\rangle_0[\mu]+\mathcal{O}(m^2)$
\cite{Maris1,fpi1} (where $m_\pi[\mu]$ is the pion mass at finite $\mu$,
$m$ is the current quark mass and $\langle\bar{q}q\rangle_0[\mu]$ is the
quark condensate in the chiral limit at finite $\mu$) one would also
conclude that $m_\pi[\mu]$ is kept unchanged from its value at $\mu=0$ when
$\mu<\mu_0$. In Ref. \cite{fpi1}, the authors did not made an analytical analysis of
$f_\pi[\mu]$ and $m_\pi[\mu]$ by the method of pole analysis, but instead made a direct numerical calculation. There exist numerical errors in this calculation. Within numerical errors $f_\pi[\mu]$ and $m_\pi[\mu]$ do not change with $\mu$ for $\mu < 300~\mathrm{MeV}$. The analytical analysis made in this paper explains the numerical results obtained in \cite{fpi1}.

For $\mu> \mu_0$, one can calculate $f_\pi[\mu]$ and $m_\pi[\mu]$ numerically based on  Eq. (\ref{fpi4}) and the Gell-Mann-Oakes-Renner relation. The behaviors of $f_\pi[\mu]$ and $m_\pi[\mu]$ for $\mu> \mu_0$ are
shown in Fig. 2 and Fig. 3. One sees that $f_\pi[\mu]$ exhibits a
sharp decrease whereas $m_\pi[\mu]$ exhibits a sharp increase near
$\mu_0$ for all three cases. This result is quite different from the result in previous literatures. For example, in a recent work \cite{Nam}, those authors also investigated $f_{\pi}$ and $m_\pi$ at finite density within the framework of the nonlocal quark model from the instanton vacuum. Their results show that in the range $0 \leq \mu \leq 320~\mathrm{MeV}$, $f_\pi$ falls slowly whereas $m_\pi$ increases slowly. This behavior of $f_\pi$ and $m_\pi$ is qualitatively different from that found in this paper.  

Finally, we should emphasize that in obtaining our results about
$f_\pi[\mu]$, $m_\pi[\mu]$ and $\langle\bar{q}q\rangle_0[\mu]$ in
this paper, we have made these approximations and assumptions: (1)
we adopt the rainbow-ladder approximation of the DSEs; (2) we assume
the quark propagator and quark-meson vertex are analytic in the
neighborhood of $\mu=0$; (3) we have neglected the $\mu$-dependence
of the dressed gluon propagator. (for a discussion about these
approximations and assumptions, see Ref. \cite{fpi1}). For further
study one should consider improvements on these approximations.
\begin{center}
\epsfig{file=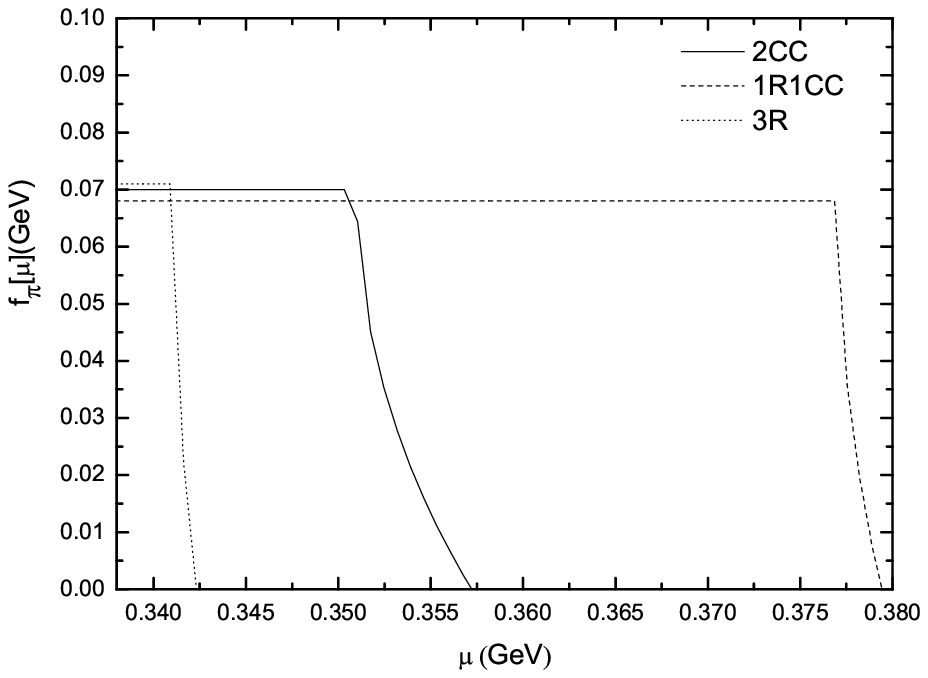, width=9cm}

\vspace{-0.4cm}

{ FIG.2. The $\mu$ dependence of $f_\pi$ near $\mu_0$.}
\end{center}

\begin{center}
\epsfig{file=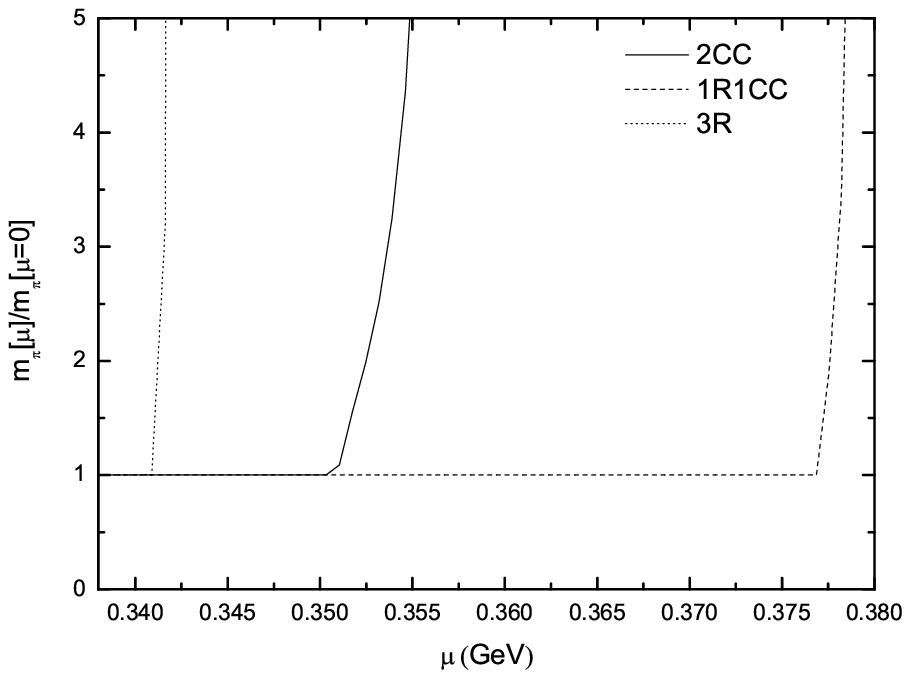, width=9cm}

\vspace{-0.4cm}

{ FIG.3. The $\mu$ dependence of $m_\pi$ near $\mu_0$.}
\end{center}

To summarize, based on the previous work in Ref. \cite{fpi1} on the quark-meson vertex and pion properties at finite quark chemical potential, we provide an analytical analysis of the weak decay constant of the pion ($f_\pi[\mu]$) and the pion mass ($m_\pi[\mu]$) at finite quark chemical potential using the model quark propagator proposed in   Ref. \cite{Alkofer04}. It is found that when $\mu$ is below a threshold value $\mu_0$ 
(which equals $0.350~\mathrm{GeV}$, $0.377~\mathrm{GeV}$ and $0.341~\mathrm{GeV}$, for the $\mathrm{2CC}$, $\mathrm{1R1CC}$ and $\mathrm{3R}$ parametrizations of the model quark propagator, respectively.), $f_\pi[\mu]$ and $m_\pi[\mu]$ are kept unchanged from their vacuum values. The value of $\mu_0$ is intimately connected with the pole distribution of the model quark propagator and is found to coincide with the threshold value below which the quark-number density vanishes identically. Numerical calculations show that when $\mu$ becomes larger than $\mu_0$, $f_\pi[\mu]$ exhibits a sharp decrease whereas $m_\pi[\mu]$ exhibits a sharp increase. These results are quite different from those obtained in previous literatures. For example, our results are qualitatively different from those reported in Ref. \cite{Nam}, which uses the nonlocal chiral quark model from the instanton vacuum to investigate $f_\pi$ and $m_\pi$ at finite density.  

\vspace*{0.4cm}

\noindent{\large \bf Acknowledgments}

This work was supported in part by the National Natural Science
Foundation of China (under Grant No 10575050) and the Research Fund
for the Doctoral Program of Higher Education (under Grant No
20060284020).

\appendix
\section{The Analysis of the Poles}
\subsection{General Analysis}

With the quark propagator given by Eq. (\ref{QuarkP1}) one can
find the following
\begin{eqnarray}
\sigma_v&=&\sum_j\left[\frac{r_j}{q^2+(a_j+ib_j)^2}
+\frac{r_j}{q^2+(a_j-ib_j)^2}\right]=\frac{f_v}{f_0}\\
\sigma_s&=&\sum_j\left[\frac{r_j(a_j+ib_j)}{q^2+(a_j+ib_j)^2}
+\frac{r_j(a_j-ib_j)}{q^2+(a_j-ib_j)^2}\right]=\frac{f_s}{f_0}
\end{eqnarray}
with
\begin{eqnarray}
f_v&=&\sum_j2r_j\left(q^2+a_j^2-b_j^2\right)\prod_{k\neq
j}[q^2+(a_k+ib_k)^2][q^2+(a_k-ib_k)^2]\label{F3}\\
f_s&=&\sum_j2r_ja_j\left(q^2+a_j^2+b_j^2\right)\prod_{k\neq
j}[q^2+(a_k+ib_k)^2][q^2+(a_k-ib_k)^2]\label{F5}\\
f_0&=&\prod_j[q^2+(a_j+ib_j)^2][q^2+(a_j-ib_j)^2].\label{F6}
\end{eqnarray}
Then one has
\begin{eqnarray}
\label{F2}F(q_4)&=&\frac{4\sigma_s} {\sigma_v^2q^2+\sigma_s^2}\left[
\sigma_s\sigma_v+2q_4^2(\sigma_s\sigma_v^\prime-
\sigma_s^\prime\sigma_v)\right]=\frac{1}{f_0}\frac{\Xi}{f_v^2q^2+f_s^2},
\end{eqnarray}
where
\begin{equation}
\Xi=4f_s\left[f_sf_v+ 2(q^2-\vec{q}^2)(f_sf_v^\prime-f_s^\prime
f_v)\right].
\end{equation}
For convenience let us use $x^2=q^2$ with $x$ a complex number. Then
the denominator of the right-hand-side of Eq. (\ref{F2}) can be
decomposed as
\begin{eqnarray}
f_v^2x^2+f_s^2&=&(f_vx+if_s)(f_vx-if_s).
\end{eqnarray}
$f_v$ and $f_s$ can be expressed as
\begin{eqnarray}
f_v&=&\sum_jr_j\left[\frac{f_0}{x^2+(a_j+ib_j)^2}+
\frac{f_0}{x^2+(a_j-ib_j)^2}\right]\\
f_s&=&\sum_jr_j\left[\frac{f_0(a_j+ib_j)}{x^2+(a_j+ib_j)^2}+
\frac{f_0(a_j-ib_j)}{x^2+(a_j-ib_j)^2}\right],
\end{eqnarray}
so one has the following
\begin{eqnarray}
&&(f_vx+if_s)(f_vx-if_s)\nonumber\\
&=&\left\{\sum_jr_jf_0\left[\frac{x+i(a_j+ib_j)}{x^2+(a_j+ib_j)^2}+
\frac{x+i(a_j-ib_j)}{x^2+(a_j-ib_j)^2}\right]\right\}\nonumber\\
&&\times\left\{\sum_jr_jf_0\left[\frac{x-i(a_j+ib_j)}{x^2+(a_j+ib_j)^2}+
\frac{x-i(a_j-ib_j)}{x^2+(a_j-ib_j)^2}\right]\right\}\nonumber\\
&=&\left\{\sum_jr_jf_0\left[\frac{1}{x-i(a_j+ib_j)}+
\frac{1}{x-i(a_j-ib_j)}\right]\right\}\nonumber\\
&&\times\left\{\sum_jr_jf_0\left[\frac{1}{x+i(a_j+ib_j)}+
\frac{1}{x+i(a_j-ib_j)}\right]\right\}.
\end{eqnarray}
$f_0$ can be expressed as
\begin{eqnarray}
f_0&=&\prod_{k_1}[x+i(a_{k_1}+ib_{k_1})]
[x+i(a_{k_1}-ib_{k_1})]\nonumber\\
&&\times\prod_{k_2}[x-i(a_{k_1}+ib_{k_1})][x-i(a_{k_1}-ib_{k_1})].
\label{app2}
\end{eqnarray}
Therefore one obtains
\begin{eqnarray}
&&\sum_jr_jf_0\left[\frac{1}{x-i(a_j+ib_j)}+
\frac{1}{x-i(a_j-ib_j)}\right]\nonumber\\
&=&\sum_j\bigg\{r_j [x-i(a_j-ib_j)+x-i(a_j+ib_j)]
\prod_{k_1}[x+i(a_{k_1}+ib_{k_1})] [x+i(a_{k_1}-ib_{k_1})]\nonumber\\
&&\times\prod_{k_2\neq
j}[x-i(a_{k_2}+ib_{k_2})][x-i(a_{k_2}-ib_{k_2})]\bigg\}\nonumber\\
&=&\prod_{k_1}[x+i(a_{k_1}+ib_{k_1})]
[x+i(a_{k_1}-ib_{k_1})]\nonumber\\
&&\times\sum_j2r_j(x-ia_j)\prod_{k\neq
j}[x-i(a_k+ib_k)][x-i(a_k-ib_k)]
\end{eqnarray}
and
\begin{eqnarray}
&&\sum_jr_jf_0\left[\frac{1}{x+i(a_j+ib_j)}+
\frac{1}{x+i(a_j-ib_j)}\right]\nonumber\\
&=&\sum_j\bigg\{r_j [x+i(a_j-ib_j)+x+i(a_j+ib_j)]
\prod_{k_1\neq j}[x+i(a_{k_1}+ib_{k_1})] [x+i(a_{k_1}-ib_{k_1})]\nonumber\\
&&\times\prod_{k_2}[x-i(a_{k_2}+ib_{k_2})][x-i(a_{k_2}-ib_{k_2})]\bigg\}\nonumber\\
&=&\prod_{k_2}[x-i(a_{k_2}+ib_{k_2})]
[x-i(a_{k_2}-ib_{k_2})]\nonumber\\
&&\times\sum_j2r_j(x+ia_j)\prod_{k\neq
j}[x+i(a_k+ib_k)][x+i(a_k-ib_k)].
\end{eqnarray}

With Eq. (\ref{app2}) one can find the following
\begin{eqnarray}
f_v^2x^2+f_s^2&=&(f_vx+if_s)(f_vx-if_s)\nonumber\\
&=&f_0\left\{\sum_j2r_j(x-ia_j)\prod_{k\neq
j}[x-i(a_k+ib_k)][x-i(a_k-ib_k)]\right\}\nonumber\\
&&\times\left\{\sum_j2r_j(x+ia_j)\prod_{k\neq
j}[x+i(a_k+ib_k)][x+i(a_k-ib_k)]\right\}.\label{function4}
\end{eqnarray}
Hence, in order to determine the poles of $F(q_4)$, one should solve the
following three equations:
\begin{eqnarray}
\label{poles1}f_0=\prod_j[x^2+(a_j+ib_j)^2][x^2+(a_j-ib_j)^2]&=&0,\\
\label{poles2}\sum_j2r_j(x-ia_j)\prod_{k\neq
j}[x-i(a_k+ib_k)][x-i(a_k-ib_k)]&=&0,\\
\label{poles3}\sum_j2r_j(x+ia_j)\prod_{k\neq
j}[x+i(a_k+ib_k)][x+i(a_k-ib_k)]&=&0.
\end{eqnarray}
Here it should be noted that if some $b_j=0$ then  $x=i a_j $
($\mathrm{or}~ x=-i a_j$)  must be the solution of Eq.
(\ref{poles2}) (or Eq. (\ref{poles3})). One should also be aware
that after finding the roots of the above equations one should
substitute them into $\Xi$ to ensure that $\Xi(x)\neq0$ (we will see
it in the discussion of 1R1CC and 3R case below). For general
$n_P$ Eq. (\ref{poles2}) (or Eq. (\ref{poles3})) is an equation of
degree $2n_P-1$ in $x$ and it is almost impossible to give the
analytic form of the solution for general $r_j$, $a_j$ and $b_j$
when $n_P\geq2$.

\subsection{Detailed Calculation of the Poles}
For 2CC case one can find the following
\begin{eqnarray}
f_v&=&q^6+d_{v1}q^4+d_{v2}q^2+d_{v3}\\
f_s&=&d_{s1}q^4+d_{s2}q^2+d_{s3}\\
f_0&=&[q^4+2(a_1^2-b_1^2)q^2+(a_1^2+b_1^2)^2]
[q^4+2(a_2^2-b_2^2)q^2+(a_2^2+b_2^2)^2],
\end{eqnarray}
where $d_{v1},d_{v2},d_{v3},d_{s1},d_{s2},d_{s3}$ are coefficients
decided by $r_j,a_j,b_j$. With parameters shown in Table I the solutions of Eq. (\ref{poles2}) and Eq. (\ref{poles3}) are found to be
$\eta_1=0.350~\mathrm{GeV},\eta_{2,3}=(0.723\pm 0.351i)~\mathrm{GeV}$.
Of course, one can directly verify 
\begin{eqnarray}
f_v^2q^2+f_s^2&=&f_0(q^2+\eta_1^2)(q^2+\eta_2^2)(q^2+\eta_3^2).
\end{eqnarray}
So the poles of $F(q_4)$ for 2CC parameters (in the upper
half complex $q_4$ plane) are 
\begin{eqnarray}
z_1&=&i\sqrt{\vec{q}^2+\eta_1^2}\,\,\,\,(\mbox{simple pole})\\
z_2&=&\chi_2+i\omega_2\,\,\,\,(\mbox{simple pole})\\
z_3&=&\chi_3+i\omega_3\,\,\,\,(\mbox{simple pole})\\
z_4&=&\chi_4+i\omega_4\,\,\,\,(\mbox{double pole})\\
z_5&=&\chi_5+i\omega_5\,\,\,\,(\mbox{double pole})\\
z_6&=&\chi_6+i\omega_6\,\,\,\,(\mbox{double pole})\\
z_7&=&\chi_7+i\omega_7\,\,\,\,(\mbox{double pole})
\end{eqnarray}
with
\begin{eqnarray}
\omega_2&=&\omega_3\nonumber\\
&=&\sqrt{\frac{\vec{q}^2+(\mbox{Re}\eta_2)^2
-(\mbox{Im}\eta_2)^2+\sqrt{[\vec{q}^2+(\mbox{Re}\eta_2)^2
-(\mbox{Im}\eta_2)^2]^2+4(\mbox{Re}\eta_2)^2(\mbox{Im}\eta_2)^2}}{2}}
\nonumber\\
\\ \chi_2&=&-\chi_3=-\frac{(\mbox{Re}\eta_2)(\mbox{Im}\eta_2)}{\omega_2}\\
\omega_4&=&\omega_5=\sqrt{\frac{\vec{q}^2+a_1^2
-b_1^2+\sqrt{(\vec{q}^2+a_1^2 -b_1^2)^2+4a_1^2b_1^2}}{2}}\\
\chi_4&=&-\chi_5=-\frac{a_1b_1}{\omega_4}\\
\omega_6&=&\omega_7=\sqrt{\frac{\vec{q}^2+a_2^2
-b_2^2+\sqrt{(\vec{q}^2+a_2^2 -b_2^2)^2+4a_2^2b_2^2}}{2}}\\
\chi_6&=&-\chi_7=-\frac{a_2b_2}{\omega_6}.
\end{eqnarray}

For 1R1CC (and 3R) case, the analysis is similar except a little
modification for correctly analyzing the degree of the poles. Because
$b_1=0$ for 1R1CC case (for 3R case, all $b_j$ equal zero) the
function $f_v$ and $f_s$ has a factor of $q^2+a_1^2$ (see Eq.
(\ref{F3}) and Eq. (\ref{F5})) which would be canceled by the same
factor in $f_0$. Therefore for 1R1CC case one should adopt the following modified
expressions
\begin{eqnarray}
f_{v1}&=&\frac{f_v}{q^2+a_1^2}\nonumber\\
&=&2r_1[q^2+(a_2+ib_2)^2][q^2+(a_2-ib_2)^2]+
r_2(q^2+a_1^2)[q^2+(a_2-ib_2)^2]\nonumber\\
&&+r_2(q^2+a_1^2)[q^2+(a_2+ib_2)^2]\\
f_{s1}&=&\frac{f_s}{q^2+a_1^2}\nonumber\\
&=&2r_1a_1[q^2+(a_2+ib_2)^2][q^2+(a_2-ib_2)^2]+
r_2(a_2+ib_2)(q^2+a_1^2)[q^2+(a_2-ib_2)^2]\nonumber\\
&&+r_2(a_2-ib_2)(q^2+a_1^2)[q^2+(a_2+ib_2)^2]\\
f_1&=&\frac{f_0}{q^2+a_1^2}\nonumber\\
&=&(q^2+a_1^2) [q^4+2(a_2^2-b_2^2)q^2+(a_2^2+b_2^2)^2].
\end{eqnarray}
According to the decomposition in Eq. (\ref{function4}) one has
\begin{eqnarray}
f_v^2q^2+f_s^2&=&f_0(q^2+\eta_1^2)(q^2+\eta_2^2)(q^2+\eta_3^2)
\end{eqnarray}
with $\eta_{1,2,3}$ being obtained by solving Eqs.
(\ref{poles2})-(\ref{poles3}). Then
\begin{eqnarray}
F(q_4)&=&\frac{4f_s[f_sf_v+2(q^2-\vec{q}^2)
(f_sf_v^\prime-f_vf_s^\prime)]}
{f_0^2(q^2+\eta_1^2)(q^2+\eta_2^2)(q^2+\eta_3^2)}\\
&=&\frac{4f_{s1}[f_{s1}f_{v1}+2(q^2-\vec{q}^2)
(f_{s1}f_{v1}^\prime-f_{v1}f_{s1}^\prime)]}
{f_1^2(q^2+\eta_1^2)(q^2+\eta_2^2)(q^2+\eta_3^2)}(q^2+a_1^2).
\end{eqnarray}
Because $\eta_1$ equal $a_1$ exactly, one would find that
$i\sqrt{\vec{q}^2+a_1^2}$ is a double pole. The analysis for 3R
case is similar.


\vspace*{0.4cm}
\begin{thebibliography}{99}
\bibitem{Delorme} J. Delorme, G. Chanfray, and M. Ericson, Nucl. Phys. A {\bf 603}, 239 (1996).
\bibitem{Kirchbach} M. Kirchbach and A. Wirzba, Nucl. Phys. A {\bf 616}, 648 (1997).
\bibitem{Kaiser} N. Kaiser and W. Weise, Phys. Lett. B {\bf 512}, 283 (2001).
\bibitem{Meissner} U.G. Meissner, J.A. Oller and A. Wirzba, Annals. Phys. {\bf 297}, 27 (2002).
\bibitem{Kim} H.C. Kim and M. Oka, Nucl. Phys. A {\bf 720}, 386 (2003).
\bibitem{Mallik} S. Mallik and S. Sarkar, Phys. Rev. C {\bf 69}, 015204 (2004).
\bibitem{Nam} S.I. Nam and H.C. Kim, Phys. Lett. B {\bf 666}, 324 (2008). 
\bibitem{Maris} P. Maris, C.D. Roberts, and S. Schmidt, Phys. Rev. C {\bf 57}, R2821 (1998).
\bibitem{Bender} A. Bender et al., Phys. Lett. B {\bf 431}, 263 (1998).
\bibitem{Bender1} A. Bender, W. Detmold, and A.W. Thomas, Phys. Lett. B {\bf 516}, 54 (2001).
\bibitem{Maris1} P. Maris, C.D. Roberts, and P.C. Tandy, P]hys. Lett. B {\bf 420}, 267 (1998).
\bibitem{DSE1} C.D. Roberts and A.G. Williams, Prog. Part. Nucl.
Phys. {\bf 33}, 477 (1994), and references therein.
\bibitem{DSE2} C.D. Roberts and S.M. Schmidt, Prog. Part. Nucl.
Phys. {\bf 45S1}, 1 (2000), and references therein.
\bibitem{DSE3} P. Maris and C.D. Roberts, Int. J. Mod Phys.
E {\bf 12}, 297 (2003).
\bibitem{DSE4} R. Alkofer and L. von Smekal, Phys. Rept.
{\bf 353}, 281 (2001); C.S. Fischer and R. Alkofer, Phys. Rev. D
{\bf 67}, 094020 (2003), and references therein.
\bibitem{fpi1} Y. Jiang, Y.M. Shi, H.T. Feng, W.M.
Sun and H.S. Zong, Phys. Rev. C {\bf 78}, 025214 (2008).
\bibitem{Zong05} H.S. Zong, L. Chang, F.Y.Hou, W.M. Sun and Y.X. Liu, Phys. Rev. C {\bf 71}, 015205 (2005).
\bibitem{Frank96} M.R. Frank and C.D. Roberts, Phys. Rev. C {\bf
53}, 390 (1996).
\bibitem{con1} Y. Jiang, Y.B. Zhang, W.M. Sun and H.S.
Zong, Phys. Rev. D {\bf 78}, 014005 (2008).
\bibitem{Alkofer04} R. Alkofer, W. Detmold, C.S. Fischer and P.
Maris, Phys. Rev. D {\bf 70}, 014014 (2004).
\bibitem{Zong08} H.S. Zong and W.M. Sun, Phys. Rev. D {\bf 78}, 054001 (2008).
\bibitem{Halasz98} M.A. Halasz, A.D. Jackson, R.E. Shrock, M.A.
Stephanov and J.J.M. Verbaarschot, Phys. Rev. D {\bf 58}, 096007
(1998).
\end{thebibliography}
\end{document}